\documentclass[12pt]{article}
\usepackage{amssymb,amsmath}
\usepackage{cite}
\newlength{\dinwidth}
\newlength{\dinmargin}
\newtheorem{theorem}{Theorem}[section]
\newtheorem{prop}[theorem]{Proposition}
\newtheorem{lemma}[theorem]{Lemma}

\newtheorem{definition}[theorem]{Definition}
 
\newenvironment{proof}{\medskip \noindent 
            {\bf Proof.}}{ \hfill $\square$ \medskip}

\newcommand{\ie}{{\it i.e.\ }}
\newcommand{\eg}{{\it e.g.\ }}
\newcommand{\etc}{{\it etc}}

\def\msc{Modular Stability Condition}
\def\idty{{\leavevmode\hbox{\rm 1\kern -.3em I}}}
\def\nind{\noindent}

\def\inet{{\{\As_i\}_{i \in I}}}
\def\irnet{{\{\Rs_i\}_{i \in I}}}

\def\As{{\cal A}}
\def\Bs{{\cal B}}

\def\Ds{{\cal D}}

\def\Hs{{\cal H}}

\def\Js{{\cal J}}

\def\Ls{{\cal L}}
\def\Ms{{\cal M}}
\def\Ns{{\cal N}}
\def\Os{{\cal O}}
\def\Ps{{\cal P}}

\def\Rs{{\cal R}}

\def\Ws{{\cal W}}

\def\Pid{{\Ps_+ ^{\uparrow}}}
\def\Lid{{\Ls_+ ^{\uparrow}}}

\def\pair{{(\Ms,\Ns)}}
\def\pairf{{(\Rs(\Os_1),\Rs(\Os_2))}}
\def\idty{{\leavevmode\hbox{\rm 1\kern -.3em I}}}

\def\nind{\noindent}
\def\RR{{\mathbb R}}
\def\CC{{\mathbb C}}

\def\C{$C^{\ast}$-}
\def\W{$W^{\ast}$-}
\def\tr{{\rm Tr}}

\def\cM{{\cal M}}
\def\cN{{\cal N}}

\def\beq{\begin{equation}}
\def\eeq{\end{equation}}
\begin{document}
\title{Yet More Ado About Nothing: \\
The Remarkable Relativistic Vacuum State\thanks{This is an expanded 
version of an invited talk
given at the Symposium "Deep Beauty: Mathematical Innovation and the 
Search for an Underlying Intelligibility of the Quantum World", held 
at Princeton University on October 3--4, 2007. }
}
\author{{\Large Stephen J.\ Summers\,}\\[5mm]
Department of Mathematics \\
University of Florida \\ Gainesville FL 32611, USA}

\date{February 19, 2009}

\maketitle 

{\abstract \noindent An overview is given of what mathematical physics
can currently say about the vacuum state for relativistic quantum
field theories on Minkowski space. Along with a review of classical
results such as the Reeh--Schlieder Theorem and its immediate and
controversial consequences, more recent results are discussed. These
include the nature of vacuum correlations and the degree of
entanglement of the vacuum, as well as the striking fact that the
modular objects determined by the vacuum state and algebras of
observables localized in certain regions of Minkowski space encode a
remarkable range of physical information, from the dynamics and
scattering behavior of the theory to the external symmetries and even
the space--time itself. In addition, an intrinsic characterization of
the vacuum state provided by modular objects is discussed.}

\section{Introduction} \label{intro}

     For millenia, the concept of nothingness, in many forms and
guises, has occupied reflective minds, who have adopted an extraordinary
range of stances towards the notion --- from holding that it is the
Godhead itself, to rejecting it vehemently as a foul blasphemy. Even
among more scientifically inclined thinkers there has
been a similar range of views \cite{Gr}. We have no intention here to
sketch this vast richness of thought about nothingness. Instead, we
shall more modestly attempt to explain what mathematical physics has
to say about nothingness in its modern scientific guise: the
relativistic vacuum state.

     What is the vacuum in modern science? Roughly speaking, it is
that which is left over after all which can possibly be removed has
been removed, where ``possibly'' refers not to ``technically
possible'' nor to ``logically possible'', but to ``physically possible''
--- that which is possible in light of (the current understanding of)
the laws of physics. The vacuum is therefore an idealization which is
only approximately realized in the laboratory and in nature. But it 
is a most useful idealization and a surprisingly rich concept.

     We shall discuss the vacuum solely in the context of the
relativistic quantum theory of systems in four spacetime dimensional
Minkowski space, although we shall briefly indicate how similar states
for quantum systems in other space--times can be defined and
studied. In a relativistic theory of systems in Minkowski space, the
vacuum should appear to be the same at every position, and in every
direction, for all inertial observers. In other words, it should
be invariant under the Poincar\'e group, the group of isometries of
Minkowski space. And since one can remove no further mass/energy from the
vacuum, it should be the lowest possible (global) energy state.
In a relativistic theory, when one removes all mass/energy, the total
energy of the resultant state is 0. 

     These {\it desiderata} of a vacuum are intuitively appealing, but
it remains to give mathematical content to these intuitions. Once this
is done, it will be seen that this state with ``nothing in it''
manifests remarkable properties, most of which have been discovered
only in the past twenty years, and many of which are {\it not} intuitively
appealing at first exposure. On the contrary, some properties of the
vacuum state have proven to be decidedly controversial.

     In order to formulate in a mathematically rigorous manner the
notion of a vacuum state and to understand its properties, it is
necessary to choose a mathematical framework which is sufficiently
general to subsume large classes of models, is powerful enough to
facilitate the proof of nontrivial assertions of physical interest,
and yet is conceptually simple enough to have a direct, if idealized,
interpretation in terms of operationally meaningful physical
quantities. Such a framework is provided by algebraic
quantum theory \cite{Em,Ar,BrRo1,BrRo2,Haag}, also called local quantum
physics, which is based on operator algebra theory, itself initially
developed by J. von Neumann for the express purpose of providing
quantum theory with a rigorous and flexible foundation \cite{Neu,NeuI}. 
This framework is briefly described in the next section, where a
rigorous definition of a vacuum state in Minkowski space is given.

     In Section \ref{first} the earliest recognized consequences of
such a definition are discussed, including such initially nonintuitive
results as the Reeh--Schlieder Theorem. Rigorous results indicating
that the vacuum is a highly entangled state are presented in Section
\ref{vaccor}. Indeed, by many measures it is a maximally entangled
state. Though some of these results have been proven quite recently,
readers who are familiar with the heuristic picture of the
relativistic vacuum as a seething broth of virtual
particle--antiparticle pairs causing wide-ranging vacuum correlations
may not be entirely surprised by their content. But there are concepts
available in algebraic quantum field theory (AQFT) which have no known
counterpart in heuristic quantum field theory, such as the
mathematical objects which arise in the modular theory of M. Tomita
and M. Takesaki \cite{Takm}, which is applicable in the setting of
AQFT. As explained in Sections \ref{cgmasec} and \ref{charact}, the
modular objects associated with the vacuum state encode a truly
astonishing amount of physical information and also serve to provide
an intrinsic characterization of the vacuum state which admits a
generalization to quantum fields on arbitrary space--times. In
addition, it is shown in Section \ref{derive} how these objects may be
used to derive the space--time itself, thereby providing, at least in
principle, a means to derive from the observables and their
preparation (the state) a space--time in which the former can be
interpreted as being localized and evolving without any {\it a priori}
input on the nature or even existence of a space--time.  We make some
concluding remarks in Section \ref{concl}.

\section{The Mathematical Framework} \label{frame}
     
     The operationally fundamental objects in a laboratory are the
preparation apparata --- devices which prepare in a repeatable manner
the individual quantum systems which are to be examined --- and the
measuring apparata --- devices which are applied to the prepared
systems and which measure the ``value'' of some observable property of
the system. The physical notion of a ``state'' can be viewed as a
certain equivalence class of such preparation devices, and the
physical notion of an ``observable'' (or ``effect'') can be
viewed as a certain equivalence class of such measuring (or
registration) devices \cite{Lud,Ar}. In principle,
therefore, these quantities are operationally determined.

     In algebraic quantum theory, such observables are represented
by self--adjoint elements of certain algebras of operators,
either \W\ or \C algebras.\footnote{Other sorts of algebras
have also been seriously considered for various reasons; see \eg 
\cite{Em,Schw,Seg}.} In this paper we shall restrict our attention primarily
to concretely represented \W algebras, which are commonly
called von Neumann algebras in honor of the person who initiated their
study \cite{NeuI}. The reader unfamiliar with
these notions may simply think of algebras $\Ms$ of bounded 
operators\footnote{Technicalities concerning topology will be 
systematically suppressed in this paper. We therefore will not discuss
the difference between \C\ and \W algebras.}
on some (separable) Hilbert space $\Hs$ (or see 
\cite{KaRi1,KaRi2,Tak1,Tak2,Tak3} for a thorough background).
We shall denote by $\Bs(\Hs)$ the algebra of all bounded operators on
$\Hs$. Physical states are represented by mathematical {\it states} $\phi$, 
\ie linear, continuous maps $\phi : \Ms \rightarrow \CC$ from the algebra 
of observables to the complex number system which take the value 1 on 
the identity map $I$ on $\Hs$ and are positive in the sense that
$\phi(A^* A) \geq 0$ for all $A \in \Ms$. An important subclass of
states consists of {\it normal states}; these are states such that
$\phi(A) = \tr(\rho A)$, $A \in \Ms$, for some {\it density matrix} 
$\rho$ acting on $\Hs$, \ie a bounded operator on $\Hs$ satisfying
the conditions $0 \leq \rho = \rho^*$ and $\tr(\rho) = 1$. A special 
case of such normal states is constituted by the {\it vector states}: 
if $\Phi \in \Hs$ is a unit vector and $P_\Phi \in \Bs(\Hs)$ is the 
orthogonal projection onto the one dimensional subspace of $\Hs$ spanned 
by $\Phi$, the corresponding vector state is given by
$$\phi(A) = \langle \Phi, A \Phi \rangle = \tr(P_\Phi A) \, , \,
A \in \Ms \, .$$
Generally speaking, theoretical physicists tacitly restrict their
attention to normal states.\footnote{It is in the context of von Neumann
algebras and normal states that classical probability theory has a
natural generalization to noncommutative probability theory; see 
\eg \cite{ReSu2}.} 

     In AQFT the spacetime localization of the observables is taken
into account. Let $\RR^4$ represent four dimensional Minkowski space
and $\Os$ denote an open subset of $\RR^4$. Since any measurement is
carried out in a finite spatial region and in a finite time, for every
observable $A$ there exist bounded regions $\Os$ containing this
``localization'' of $A$.\footnote{It is clear from operational 
considerations that one could not expect to determine a minimal 
localization region for a given observable experimentally. In \cite{Ku3}
the possibility of determining such a minimal localization region 
in the idealized context of AQFT is discussed at length. However, the
existence of such a region is not necessary for any results in AQFT
known to the author.} We say that the observable $A$ is localized in 
any such region $\Os$ and denote by $\Rs(\Os)$ the von Neumann algebra
generated by all observables localized in $\Os$.  Clearly, it follows
that if $\Os_1 \subset \Os_2$, then $\Rs(\Os_1) \subset
\Rs(\Os_2)$. This yields a net $\Os \mapsto \Rs(\Os)$ of observable
algebras associated with the experiment(s) in question. In turn, this
net determines the smallest von Neumann algebra $\Rs$ on $\Hs$
containing all $\Rs(\Os)$.  The preparation procedures in the
experiment(s) then determine states $\phi$ on $\Rs$, the global 
observable algebra.

     Given a state $\phi$ on $\Rs$, one can construct \cite{Em,KaRi1,Tak1} 
a Hilbert space $\Hs_{\phi}$, a distinguished unit vector 
$\Omega_\phi \in \Hs_{\phi}$ and a (\C)homomorphism 
$\pi_{\phi}\colon \Rs \to \Bs(\Hs_{\phi})$, so that 
$\pi_{\phi}(\Rs)$ is a (\C)algebra acting on the Hilbert space 
$\Hs_{\phi}$, the set of vectors 
$\pi_{\phi}(\Rs)\Omega_\phi = \{ \pi_\phi(A)\Omega_\phi \mid A \in \Rs \}$ 
is dense in $\Hs_\phi$ and
$$\phi(A) = \langle \Omega_\phi, \pi_\phi(A) \Omega_\phi \rangle \, ,
\, A \in \Rs \, .$$
The triple $(\Hs_\phi,\Omega_\phi,\pi_\phi)$ is
uniquely determined up to unitary equivalence by these properties, and
$\pi_{\phi}$ is called the GNS representation of $\Rs$ determined by
$\phi$. Only if $\phi$ is a normal state is $\pi_{\phi}(\Rs)$ a
von Neumann algebra and can $(\Hs_\phi,\Omega_\phi,\pi_\phi)$ be
identified with (a subrepresentation of) $(\Hs,\Omega,\Rs)$ such that
$\Omega_\phi \in \Hs$.\footnote{If the state $\phi$ is not normal, then
the state vectors in $\Hs_\phi$ are, in a certain mathematically rigorous
sense, {\it orthogonal} to those in $\Hs$. The vacuum state of a fully
interacting model, as opposed to an interacting theory with various kinds of
cutoffs introduced precisely so that it may be realized in Fock space, 
is not normal with respect to Fock space, which is the
representation space for the corresponding free theory --- see \eg 
\cite{GlJa,BLT}. For further perspective on this issue, see \cite{SuvN}.} 
Hence, a state determines a concrete, though idealized, representation of 
the experimental setting in a Hilbert space.

     In this setting, relativistic covariance is expressed through
the presence of a representation 
$\Pid \ni \lambda \mapsto \alpha_\lambda$ of the identity component
$\Pid$ of the Poincar\'e group by automorphisms 
$\alpha_\lambda : \Rs \rightarrow \Rs$ of $\Rs$ such that
$$\alpha_\lambda(\Rs(\Os)) = \Rs(\lambda \Os) \, , $$
for all $\Os$ and $\lambda$, where 
$\alpha_\lambda(\Rs(\Os)) = \{ \alpha_\lambda(A) \mid A \in \Rs(\Os) \}$
and \newline
$\lambda \Os = \{ \lambda(x) \mid x \in \Os \}$. One says that
a state $\phi$ is {\it Poincar\'e invariant} if 
$\phi(\alpha_\lambda(A)) = \phi(A)$ for all $A \in \Rs$ and 
$\lambda \in \Pid$. In this case, there then exists a unitary representation
$\Pid \ni \lambda \mapsto U_\phi(\lambda)$ acting on $\Hs_\phi$,
leaving $\Omega_\phi$ invariant, and implementing the action of the 
Poincar\'e group:
$$U_\phi(\lambda) \pi_\phi(A) U_\phi(\lambda)^{-1} = 
\pi_\phi(\alpha_\lambda(A)) \, , $$
for all $A$ and $\lambda$. If the joint spectrum of the self--adjoint
generators of the translation subgroup $U_\phi(\RR^4)$ is contained in
the forward lightcone, then $U_\phi(\Pid)$ is said to satisfy the
(relativistic) {\it spectrum condition}. This condition is a
relativistically invariant way of requiring that the total energy in
the theory be nonnegative with respect to every inertial frame of
reference and that the quantum system is stable in the sense that it
cannot decay to energies below that of the vacuum state.

     We can now present a standard rigorous definition of a vacuum
state, which incorporates all of the intuitive {\it desiderata} discussed
above.

\begin{definition} \label{vacuum}
A vacuum state is a Poincar\'e invariant state $\phi$ on $\Rs$ such that 
$U_\phi(\Pid)$ satisfies the spectrum condition.\footnote{Some 
authors just require of a vacuum state that it be invariant under
the translation group and satisfy the spectrum condition. For the 
purposes of this paper, it is convenient to adopt the more
restrictive of the two standard definitions.} The corresponding GNS
representation $(\Hs_\phi,\Omega_\phi,\pi_\phi)$ is called a vacuum 
representation of the net of observable algebras.
\end{definition}

\nind  Note that after choosing an inertial frame of reference,
the self--adjoint generator $H$ of the time translation subgroup 
$U_\phi(t)$, $t \in \RR$, carries the interpretation of the total
energy operator and that, by definition, $H \Omega_\phi = 0$,
if $\phi$ is a vacuum state. Moreover, the total momentum operator
$\vec{P}$ and the total mass operator $M \equiv \sqrt{H^2 - \vec{P}^2} \geq 0$ 
also annihilate the vacuum ($M \Omega_\phi = 0 = \vec{P} \Omega_\phi$).

     Such vacuum states, and hence such vacuum representations,
actually exist. In the case of four dimensional Minkowski space,
vacuum representations for quantum field models with trivial
$S$-matrix have been rigorously constructed by various means (cf. \eg
\cite{WiGa,Ar2,BLT,GlJa2,BrGuLo3}) and, more recently, the same has been
accomplished for quantum field models with nontrivial scattering
matrices \cite{GrLe,BuSu2,BS6}. For two, resp. three, dimensional
Minkowski space, fully interacting quantum field models in vacuum
representations have been constructed, cf. \eg
\cite{GlJa,GlJa2,Le3,BLT}. Moreover, general conditions are known under
which to a quantum field model without a vacuum state can be (under
certain conditions uniquely) associated a vacuum representation which
is physically equivalent and locally unitarily equivalent to it
\cite{BuFr,BuWa,Dy}. Hence, the mathematical existence of a vacuum
state is often assured even in models which are not initially provided
with one.

     It will be useful in the following to describe two special
classes of spacetime regions in Minkowski space. A double cone is a 
(nonempty) intersection of an open forward lightcone with an open backward
lightcone. Such regions are bounded, and the set $\Ds$ of all double
cones is left invariant by the natural action of $\Pid$ upon it. An
important class of unbounded regions is specified as follows.
After choosing an inertial frame of reference, one defines the right wedge
to be the set \newline
$W_R = \{ x = (t,x_1,x_2,x_3) \in \RR^4 \mid x_1 > \vert t \vert \}$
and the set of wedges to be \newline
$\Ws = \{ \lambda W_R \mid \lambda \in \Pid \}$.
The set of wedges is independent of the choice of reference frame;
only which wedge is designated the right wedge is frame-dependent.

\section{Immediate Consequences} \label{first}

     We now turn to some immediate consequences of the definition of a
vacuum state. One of the most controversial was also one of the first
to be noted. In order to avoid a too heavily laden notation, and since
in this and the next section our starting point is a vacuum
representation, we shall drop the subscript $\phi$ and the symbol
$\pi_\phi$ (\ie we identify $\Rs(\Os)$ and $\pi_\phi(\Rs(\Os))$). A 
vacuum representation is said to satisfy {\it weak additivity} if for 
each nonempty $\Os$ the smallest von Neumann algebra containing
$$\{ U(x) \Rs(\Os) U(x)^{-1} \mid x \in \RR^4 \} $$
coincides with $\Rs$. This is a weak technical assumption satisfied in 
most models; for example, it holds in any theory in which there is a 
Wightman field locally associated with the observable algebras 
(see, \eg \cite{BW1,BW2,DrSuWi}). 

     Let $\Os$ be an open subset of $\RR^4$ and let $\Os'$ denote
the interior of its causal complement, the set of all points 
in $\RR^4$ which are spacelike separated from all points in $\Os$.
A net $\Os \mapsto \Rs(\Os)$ is said to be {\it local} (or to satisfy
{\it locality}) if whenever
$\Os_1 \subset \Os_2{}'$ one has $\Rs(\Os_1) \subset \Rs(\Os_2)'$,
where $\Rs(\Os)'$, the commutant of $\Rs(\Os)$, represents the set
of all bounded operators on $\Hs$ which commute with all elements
of $\Rs(\Os)$. Ordinarily, this property of locality is viewed
as a manifestation of Einstein causality, which posits that signals 
and causal influences cannot propagate faster than the speed of light,
and therefore spacelike separated quantum systems must be independent
in some sense. As is the case with so many received notions, there
is much more here than meets the eye initially; but this is not
the place to address this matter (cf. \cite{But,Su2,Susub} for certain 
aspects of this point). We wish to emphasize that locality
will not be a standing assumption in this paper. If a net is
not explicitly assumed to be local, then the property is not necessary
for the respective result. And, in fact, locality will be {\it derived}
in the settings discussed in Sections \ref{charact} and 
\ref{derive}.\footnote{For a very different derivation of locality, 
see \cite{BS5}.} 

     For vacuum representations of local nets in which weak additivity
is satisfied, the Reeh--Schlieder Theorem holds (cf. 
\cite{BLT,Haag,Ar,Jo,StWi}).

\begin{theorem} \label{R-S} 
Consider a vacuum representation of a local net fulfilling the condition
of weak additivity. For every nonempty region $\Os$ such that 
$\Os' \neq \emptyset$, the vector $\Omega$ is cyclic and separating for 
$\Rs(\Os)$, \ie the set of vectors $\Rs(\Os)\Omega$ is dense in $\Hs$, resp. 
$A \in \Rs(\Os)$ and $A\Omega = 0$ entail $A = 0$.\footnote{Note that
even if the net of observable algebras is not local, $\Omega$ is still
cyclic for $\Rs(\Os)$.} 
\end{theorem}

    There are two distinct aspects to this theorem. First of all, the
fact that the vacuum is separating for local observables means exactly
that no nonzero local observable can annihilate $\Omega$. Hence,
any event represented by a nonzero projection $P \in \Rs(\Os)$
must have nonzero expectation in the vacuum state: 
$\langle \Omega, P \Omega \rangle > 0$. In the vacuum, any local 
event can occur! Moreover, $0 < C = C^* \in \Rs(\Os)$ entails the 
existence of an element $0 \neq A \in \Rs(\Os)$ such that $C = A^*A$; thus 
$\langle \Omega, C \Omega \rangle = \Vert A \Omega \Vert^2$, 
which also yields $\langle \Omega, C \Omega \rangle > 0$ in this more
general case. Therefore the stress--energy density tensor $T(x)$
smeared with any test function with compact support cannot be a
positive operator in a vacuum representation \cite{EpGlJa} (in fact,
it is unbounded below), in contrast to the situation in classical
physics, since its vacuum expectation is zero. Furthermore, in light of 
the fact that the vacuum state 
contains no real particles ($M \Omega = 0$), it follows that there can be no
localized particle counters. Indeed, if $C \in \Rs$ is a particle counter 
for a particle described in the model, then $C = C^* > 0$ and 
$\langle \Omega, C \Omega \rangle = 0$. Therefore, $C$ cannot be an 
element of any algebra $\Rs(\Os)$ with $\Os'$ nonempty. 
Hence the notion of particle in relativistic quantum field theory cannot 
be quite as simple as classical mechanics would have it. It has even 
been argued that the notion is nonsensical in relativistic quantum field 
theory, but this is not the place for further discussion of this point,
either. (See, however, \cite{Haag,HaCl2,Sch0,BPS,Bu,Fr}.)

     Second, there is the cyclicity of the vacuum for all local
algebras: every vector state in the vacuum representation can be
arbitrarily well approximated using vectors of the form $A \Omega$, $A
\in \Rs(\Os)$, no matter how small in extent $\Os$ may be.  Hence, the
class of all states resulting from the action of arbitrary operations
upon the vacuum is effectively indistinguishable from the class of
states resulting from operations performed in arbitrarily small
spacetime regions upon the vacuum. {\it Prima facie}, such a state
would seem to be different from the vacuum only in a region which one
can make as small as one desires. In our view, a reasonable physical
picture of this situation is indicated in this way: an experimenter
in any given region $\Os$ can, in principle, perform measurements
designed to exploit nonlocal vacuum fluctuations (see the next
section) in such a manner that any prescribed state can be reproduced
with any given accuracy.  These consequences of cyclicity also
unleashed some controversy, some of which is well discussed in
\cite{Ha} (see also \cite{Red}). We shall not elaborate upon these
matters here, except to point out the fact that the existing proposals
to avoid Reeh--Schlieder by changing the notion of localization (1)
are necessarily restricted to free quantum field models and (2)
introduce at least as many problems as they ``solve'', see \eg
\cite{Ha}.

     We wish to emphasize that these (for some readers disturbing)
properties are by no means unique to the vacuum --- the
Reeh--Schlieder Theorem is valid for {\it any} vector in the vacuum
representation which is analytic for the energy \cite{Bor1}; in
particular, it holds for any vector with finite energy content. So its
conclusions and various consequences are true of {\it all} physically
realizable vector states in the vacuum representation, since any
preparation can only implement a finite exchange of energy!

\section{Vacuum Correlations} \label{vaccor}

     We turn to what is rigorously known about the nature of vacuum
correlations, preparing first some definitions to be used in this
section. Given a pair $\pair$ of algebras representing the observable
algebras of two subsystems of a given quantum system, a state $\phi$
is said to be a {\it product state} across $\pair$ if $\phi(MN) =
\phi(M)\phi(N)$ for all $M \in \Ms$, $N \in \Ns$.  In such states, the
observables of the two subsystems are not correlated and the
subsystems manifest a certain kind of independence --- see \eg
\cite{Su2}. A normal state $\phi$ on $\Ms \bigvee \Ns$ is {\it
separable}\footnote{also termed decomposable, classically correlated,
or unentangled by various authors} if it is in the norm closure of the
convex hull of the normal product states across $\pair$, \ie it is a
mixture of normal product states.  Otherwise, $\phi$ is said to be
{\it entangled} (across $(\Ms,\Ns)$).\footnote{This terminology is
becoming standard in quantum information theory, but there are still
physicists who tacitly restrict their attention to vector states on
mutually commuting algebras of observables which are isomorphic to
full matrix algebras, \ie they consider only {\it pure} states, which
are entangled if and only if they are not product states.} From the
point of view of what is now called quantum information theory, the
primary difference between classical and quantum theory is the
existence of entangled states in quantum theory. In fact, only if both
subsystems are quantum, \ie both algebras are noncommutative, do there
exist entangled states on the composite system \cite{Rag}. Although
not understood at that time in this manner, some of the founders of
quantum theory realized as early as 1935 \cite{EPR,Schr} that such
entangled states were the source of the ``paradoxical'' behavior of
quantum theory (as viewed from the vantage point of classical
physics). Today, entangled states are regarded as a resource to be
employed in order to carry out tasks which cannot be done classically,
\ie only with separable states --- cf. \cite{Hor,Key,WW}.

     Another direct consequence of the Reeh--Schlieder Theorem is that
for all nonempty spacelike separated $\Os_1,\Os_2$ with nonempty 
causal complements, no matter how far spacelike separated they may be, 
there exist many projections $P_i \in \Rs(\Os_i)$ which are positively 
correlated in the vacuum state, \ie such that 
$\phi(P_1 P_2) > \phi(P_1) \phi(P_2)$.

\begin{theorem}  \label{mine}
Consider a vacuum representation of a local net fulfilling the condition
of weak additivity, and let $\Os_1,\Os_2$ be any nonempty spacelike separated 
regions with nonempty causal complements. Let $\phi$ be any state induced
by a vector analytic for the energy (\eg the vacuum state). Then for 
any projection $P_1 \in \Rs(\Os_1)$ with $0 \neq P_1 \neq I$ 
there exists a projection 
$P_2 \in \Rs(\Os_2)$ such that $\phi(P_1 P_2) > \phi(P_1) \phi(P_2)$. 
\end{theorem}

     This is an immediate consequence of Theorem \ref{R-S} and the
following lemma, the proof of which is implicit in the proof of
Theorem 5 in \cite{Red}. For the convenience of the reader, we make
this explicit here.

\begin{lemma}
Let $\Ms$ and $\Ns$ be von Neumann algebras on $\Hs$ with $\Omega \in \Hs$
a unit vector cyclic for $\Ns$ and separating for $\Ms$, and let $\omega$ 
be the corresponding state induced upon $\Bs(\Hs)$. Then for any projection
$P \in \Ms$ with $0 \neq P \neq I$, there exists a projection $Q \in \Ns$
such that $\omega(PQ) > \omega(P)\omega(Q)$.
\end{lemma}

\begin{proof}
Let $P \in \Ms$ be a projection with $0 \neq P \neq I$. It suffices 
to establish the existence of a projection $Q \in \Ns$ such that 
$\omega(PQ) \neq \omega(P)\omega(Q)$, since, if necessary, $Q$ can
be replaced by $I - Q \in \Ns$ to yield the assertion. So assume
for the sake of contradiction that
$\omega(PQ) = \omega(P)\omega(Q)$, for all such $Q$. Then with
$\widehat{P} = P - \omega(P) \cdot I \in \Ms$, one has
$\omega(\widehat{P}Q) = 0$, for all projections $Q \in \Ns$. By the 
spectral theorem, this entails $\omega(\widehat{P}N) = 0$, 
for all $N \in \Ns$, \ie
$$\langle \widehat{P} \Omega, N \Omega \rangle = 0 \, , \, N \in \Ns \, .$$
Since $\Omega$ is cyclic for $\Ns$, this yields $\widehat{P} \Omega = 0$, 
so that $\widehat{P} = 0$, \ie $P = \omega(P) \cdot I$. Since $P = P^2$,
this entails $\Vert P\Omega \Vert^2 = \langle P\Omega, P\Omega\rangle = 
\omega(P) \in \{ 0,1 \}$, 
\ie either $P\Omega = 0$ or $P\Omega = \Omega$. Since $\Omega$ is
separating for $\Ms$, this implies either $P = 0$ or $P = I$ holds, 
a contradiction in either case.
\end{proof}

     The fact that vacuum fluctuations enable such generic
``superluminal correlations'' has also generated controversy, since
they seem to challenge received notions of causality. This is another
complex matter which we cannot go into here, but at least some forms of 
causality have been proven in AQFT (for recent discussions, see \eg 
\cite{ReSu1,But}) and therefore are completely compatible with such 
correlations.

     Of course, Theorem \ref{mine} entails that the vacuum is not a
product state across $\pairf$, but not yet that it is entangled across
$\pairf$. Much finer analyses of the nature and degree of the
entanglement of the vacuum state have been carried out in the
literature, and we shall explain some of these. A quantitative measure
of entanglement is provided by using {\it Bell correlations}.  The
following definition was made in \cite{SW1}.

\begin{definition} Let $\Ms,\Ns \subset \Bs(\Hs)$ be von Neumann algebras
such that $\Ms \subset \Ns{}'$. The maximal Bell correlation of
the pair $(\cM,\cN)$ in the state $\phi$ is
$$\beta(\phi,\cM,\cN) \equiv \sup \, \frac{1}{2} \,
\phi(M_1(N_1+N_2)+M_2(N_1-N_2))\, , $$
where the supremum is taken over all self-adjoint
$M_i \in \cM, N_j \in \cN$ with norm less than or equal to 1.
\end{definition}

     As explained in \eg \cite{SW2}, the CHSH version
of Bell's inequalities can be formulated in algebraic quantum
theory as
\begin{equation}  \label{bell}
\beta(\phi,\cM,\cN) \leq 1 \, .
\end{equation}
If $\phi$ is separable across $\pair$, then $\beta(\phi,\Ms,\Ns) = 1$
\cite{SW2}. Hence states which violate Bell's inequalities are necessarily
entangled, though the converse is not true (cf. \cite{WW} for a
discussion and references). Whenever at least one of the systems is 
classical, the bound (\ref{bell}) is satisfied in {\it every} state:

\begin{prop} [\cite{SW2}]
Let $\Ms,\Ns \subset \Bs(\Hs)$ be mutually commuting von Neumann algebras.
If either $\Ms$ or $\Ns$ is abelian, then $\beta(\phi,\Ms,\Ns) = 1$
for all states $\phi$ on $\Bs(\Hs)$.
\end{prop}

     If, on the other hand, both algebras are nonabelian, then there
always exists a state in which the inequality (\ref{bell}) is (maximally) 
violated, as long as the Schlieder property holds, \ie $MN = 0$ for
$M \in \Ms$ and $N \in \Ns$ entail either $M=0$ or $N = 0$ \cite{Lan}. 
Because it is known \cite{Cir,SW2} that 
$1 \leq \beta(\phi,\cM,\cN) \leq \sqrt{2}$, for all states $\phi$ on 
$\Bs(\Hs)$, one says that if
$\beta(\phi,\cM,\cN) = \sqrt{2}$, then the pair $(\Ms,\Ns)$ maximally
violates Bell's inequalities in the state $\phi$.

     In \cite{SW5} it is shown under quite general physical assumptions 
that in a vacuum representation of a local net one has 
$\beta(\phi,\Rs(W),\Rs(W')) = \sqrt{2}$, for every wedge $W$ and 
{\it every} normal state $\phi$. In particular, Bell's inequalities 
are {\it maximally} violated in the vacuum state. In addition,
under somewhat more restrictive but still general assumptions which include
free quantum field theories and other physically relevant models, it is
shown in \cite{SW5} that 
$\beta(\phi,\Rs(\Os_1),\Rs(\Os_2)) = \sqrt{2}$, for
any two spacelike separated double cones whose closures intersect
(\ie tangent double cones) and {\it all} normal states $\phi$. Hence, such
pairs of observable algebras also maximally violate Bell's inequalities
in the vacuum.
     
     Commonly, physicists say that theories violating Bell's
inequalities are ``nonlocal''; yet, here are fully local models
maximally violating Bell's inequalities. This linguistic confusion is
probably so profoundly established by usage that it cannot be
repaired, but the reader should be aware of the distinct meanings of
these two uses of ``local''. The former refers to nonlocalities in
certain correlations (in certain states), while the latter refers to
the commensurability of observables localized in spacelike separated
spacetime regions. So the former is a property of states, while
the latter is a property of observable algebras.  The results
discussed above establish the generic compatibility of the former sort
of ``nonlocality'' with the latter kind of ``locality''. The wary
reader should always ascertain which sense of ``local'' is being
employed by a given author.

     In the now quite extensive quantum information theory literature,
there are various attempts to quantify the degree of entanglement of a
given state (cf. \eg \cite{Hor,Key}), but these agree that maximal
violation of inequality (\ref{bell}) entails maximal
entanglement. Thus, the vacuum state is maximally entangled and
thereby describes a maximally non-classical situation.

     The localization regions for the observable algebras which have
been proven to manifest maximal violation of Bell's inequality in the
vacuum (indeed, in every state) are spacelike separated but
tangent. If the double cones have nonzero spacelike separation, any
violation of Bell's inequality in the vacuum cannot be maximal:

\begin{prop} [\cite{SW6,SW1,SW2}] Let $\Os \mapsto \Rs(\Os)$ be a local
net in an irreducible vacuum representation with a lowest mass
$m > 0$. Then for any pair $(\Os_1,\Os_2)$ of spacelike separated
regions one has
$$\beta(\phi,\Rs(\Os_1),\Rs(\Os_2)) \leq \sqrt{2} - 
\frac{\sqrt{2}}{7 + 4\sqrt{2}}(1 - e^{-md(\Os_1,\Os_2)}) $$
(optimal for smaller $d(\Os_1,\Os_2)$) and
$$\beta(\phi,\Rs(\Os_1),\Rs(\Os_2)) \leq 1 + 2 e^{-md(\Os_1,\Os_2)} $$
(optimal for larger $d(\Os_1,\Os_2)$), where $\phi$ is a vacuum state 
and $d(\Os_1,\Os_2)$ is the maximal timelike distance $\Os_1$ can be 
translated before it is no longer spacelike separated from $\Os_2$.
\end{prop}

      Hence, if $d(\Os_1,\Os_2)$ is much larger than a few Compton
wavelengths of the lightest particle in the theory, then any violation
of Bell's inequality in the vacuum would be too small to be
observed. As explained in \cite{SW2}, if there are massless particles in
the theory, then the best decay in the vacuum Bell correlation one can
expect is proportional to $d(\Os_1,\Os_2)^{-2}$. Although the decay in
the massless case is much weaker, experimental apparata have nonzero
lower bounds on the particle energies they can effectively
measure. Such nonzero sensitivity limits would serve as an effective
lowest mass, leading to an exponential decay once again
\cite{SW2}. Nonetheless, attempts have been made to obtain lower
bounds on the Bell correlation $\beta(\phi,\Rs(\Os_1),\Rs(\Os_2))$ as
a function of $d(\Os_1,\Os_2)$. As the published results have only
treated some very special models and very special observables, we
shall refrain from discussing these here (but cf. \cite{ReReSi} and
references given there).

     Nonetheless, using properties of $\beta(\phi,\Ms,\Ns)$
established by the author and R.F. Werner \cite{SW6}, H. Halvorson
and R. Clifton have proven the following result, which entails that in
a vacuum representation in which weak additivity and locality hold,
the vacuum state (and any state induced by a vector analytic for the
energy) is entangled across $\pairf$ for arbitrary nonempty spacelike
separated regions $\Os_1,\Os_2$.

\begin{theorem} [\cite{HaCl1}]  \label{HC}
Let $\Ms$ and $\Ns$ be nonabelian von Neumann algebras acting on $\Hs$ 
such that $\Ms \subset \Ns'$. If $\Omega \in \Hs$ is cyclic for
$\Ms$ and $\omega$ is the state on $\Bs(\Hs)$ induced by $\Omega$,
then $\omega$ is entangled across $(\Ms,\Ns)$. 
\end{theorem}

     The proof does not provide a lower bound on
$\beta(\phi,\Ms,\Ns)$.  For further discussion and references
concerning the violation of Bell's inequalities in algebraic quantum
theory, see \cite{Su3,SW6,Red,HaCl1}.

     Though model independent lower bounds on
$\beta(\phi,\Rs(\Os_1),\Rs(\Os_2))$ are not yet available, R. Verch
and Werner \cite {VeWe} have obtained model independent results
on the nature of the entanglement of the vacuum state across
nontangent pairs $\pairf$ in terms of some further notions currently
employed in quantum information theory, which go beyond Theorem
\ref{HC}. They proposed the following definition \cite{VeWe}.

\begin{definition} 
Let $\Ms$ and $\Ns$ be von Neumann algebras acting upon a Hilbert 
space $\Hs$. A state $\phi$ on $\Bs(\Hs)$ has the \textnormal{ppt property} 
if for any choice of finitely many $M_1,\ldots,M_k \in \Ms$ and 
$N_1,\ldots,N_k \in \Ns$, one has 
$$\sum_{\alpha,\beta} \phi(M_\beta M_\alpha^* N_\alpha^* N_\beta) \geq 0
\, .$$ 
\end{definition}

\nind They show that this generalizes the notion of states with
positive partial transpose familiar from quantum information theory
\cite{Pe}, a notion restricted to finite dimensional Hilbert spaces
prior to \cite{VeWe}.  They also show that if a state is ppt, then it
satisfies Bell's inequalities, and they prove that any separable state is
ppt. Indeed, in general the class of ppt states properly contains the
class of separable states.

     Another notion from quantum information theory is that of
distillability (of entanglement). Roughly speaking, this refers to
being able to operate upon a given state in certain (local) ways to
increase its entanglement across two subsystems.  Separable states are
not distillable; they are not entangled, and operating upon them in
the allowable manner will not result in an entangled state.  We refer
the reader to \cite{VeWe} for a discussion of the general case and
restrict ourselves here to a discussion of the following special case.

\begin{definition} [\cite{VeWe}]
Let $\Ms$ and $\Ns$ be von Neumann algebras acting upon a Hilbert 
space $\Hs$. A state $\phi$ on $\Bs(\Hs)$ is \textnormal{1-distillable} 
if there exist completely positive maps $T : \Bs(\CC^2) \rightarrow \Ms$
and $S : \Bs(\CC^2) \rightarrow \Ns$ such that the functional
$\omega(X \otimes Y) \equiv \phi(T(X)S(Y))$, 
$X \otimes Y \in \Bs(\CC^2) \otimes \Bs(\CC^2)$ is not ppt.
\end{definition}

     Verch and Werner show that 1-distillable states are distillable
and not ppt. They also prove the following theorem.

\begin{prop} [\cite{VeWe}]
Let $\Os \mapsto \Rs(\Os)$ be a local net in a vacuum 
representation satisfying weak additivity. Then if $\Os_1$ and
$\Os_2$ are strictly spacelike separated double cones, the vacuum
state is 1-distillable across the pair $(\Rs(\Os_1),\Rs(\Os_2))$.
\end{prop}

\nind Hence, the vacuum is distillable and not ppt across $\pairf$ no
matter how large $d(\Os_1,\Os_2)$ is. We remark that, once again, this
theorem is valid also for states induced by vectors in the vacuum 
representation which are analytic for the energy \cite{VeWe}. For a 
discussion of some further aspects of the entanglement of the vacuum in AQFT, 
we refer the reader to \cite{ClHa}.

\section{Geometric Modular Action} \label{cgmasec}

     We emphasize that nearly all of the remarkable properties of the
vacuum state discussed to this point are shared by all vector states
which are analytic for the energy. In the remainder of this paper we shall
be dealing with properties unique to the vacuum.

     A crucial breakthrough in the theory of operator algebras was the
Tomita--Takesaki theory \cite{Takm} (see also \cite{KaRi2,Tak2}), which
is proving itself to be equally powerful and productive for the
purposes of mathematical quantum theory.
One of the settings subsumed by this theory is a von Neumann algebra
$\Ms$ with a cyclic and separating vector $\Omega \in \Hs$. The
data $(\Ms,\Omega)$ then uniquely determine an antiunitary 
involution\footnote{commonly called the modular conjugation or 
modular involution associated with $(\Ms,\Omega)$}
$J \in \Bs(\Hs)$ and a strongly continuous group of unitaries
$\Delta^{it}$, $t \in \RR$,\footnote{$\Delta$ is a certain, typically 
unbounded, positive operator called the modular operator associated 
with $(\Ms,\Omega)$} 
such that $J \Omega = \Omega = \Delta^{it} \Omega$, $J \Ms J = \Ms'$ and
$ \Delta^{it} \Ms \Delta^{-it} = \Ms$, for all $t \in \RR$,
along with further significant properties. From the Reeh--Schlieder
Theorem (Theorem \ref{R-S}), this theory is applicable to the pair
$(\Rs(\Os), \Omega)$, under the indicated conditions. Since, as explained
above, the algebras and states are operationally determined (in
principle), the corresponding modular objects $J_\Os, \Delta_\Os^{it}$
are, as well.

     In pathbreaking work \cite{BW1,BW2}, J.J. Bisognano and
E.H. Wichmann showed that for a net of von Neumann algebras 
$\Os \mapsto \Rs(\Os)$ locally associated with a finite--component 
quantum field satisfying the Wightman axioms \cite{Jo,StWi,BLT} 
(and therefore in a vacuum representation), the modular objects 
$J_W, \Delta_W^{it}$ determined by the wedge algebras 
$\Rs(W)$, $W \in \Ws$, and the vacuum vector $\Omega$ have a geometric 
interpretation\footnote{See also \cite{DrSuWi} for later advances in 
this particular setting.}:
\begin{equation} \label{modcov}
\Delta_W^{it} = U(\lambda_W(2\pi t)) \, , 
\end{equation}
for all $t \in \RR$ and $W \in \Ws$, where 
$\{\lambda_W(2\pi t) \mid t \in\RR\} \subset \Pid$ is the one-parameter 
subgroup of boosts leaving $W$ invariant. Explicitly for $W = W_R$,
$$ \lambda_{W_R}(t) =  
              \left( \begin{array}{cccc} \cosh t & \sinh t & 0 & 0 \\ 
                     \sinh t & \cosh t & 0 & 0 \\
                     0 & 0 & 1 & 0 \\
                     0 & 0 & 0 & 1  \end{array} \right)   \, .  $$
The relation (\ref{modcov}) has come to be referred to as {\it modular
covariance}. Moreover, for scalar Boson fields\footnote{See \cite{BW2} 
for arbitrary finite-component Wightman fields.}, one has
\begin{equation} \label{Jtheta}
J_{W_R} = \Theta U_{\pi} \, ,
\end{equation}
where $\Theta$ is the PCT-operator associated to the Wightman field
and $U_{\pi}$ implements the rotation through the angle $\pi$ about 
the $1$-axis, with similar results for general wedge $W \in \Ws$.
Hence, one has
\begin{equation} \label{Jcgma}
J_{W_R} \Rs(\Os) J_{W_R} = \Rs(\theta_R \Os) \, ,
\end{equation}
for all $\Os$, where $\theta_R \in \Ps_+$ is the reflection through 
the edge \newline
$\{ (0,0,x_2,x_3) \mid x_2,x_3 \in \RR \}$ of the wedge $W_R$. This
implies in turn that for all $W \in \Ws$ one has
\begin{equation} \label{BWcgma}
J_{W} \, \{ \Rs(\widetilde{W}) \mid \widetilde{W} \in \Ws \} \, J_{W} = 
\{ \Rs(\widetilde{W}) \mid \widetilde{W} \in \Ws \} \, .
\end{equation}
Thus the adjoint action of the modular involutions $J_W$, $W \in \Ws$, 
leaves the set \newline
$\{ \Rs(W) \mid W \in \Ws \}$ of observable algebras associated
with wedges invariant, \ie wedge algebras are transformed to wedge
algebras by this adjoint action.

     Although in the special case of the massless free scalar field
\cite{HiLo} (and, more generally, for conformally invariant quantum
field theories \cite{BrGuLo1}) also the modular objects corresponding
to $(\Rs(\Os),\Omega)$ for $\Os \in \Ds$ have geometric meaning, some
explicit computations in the free massive field have indicated that
this is not true in general. Moreover, as we shall see in the next
section, {\it only} the vacuum vector $\Omega$ yields modular objects
having any geometric content. This fact yields an intrinsic
characterization of the vacuum state.

     But before we explore this noteworthy state of affairs, let us
examine some of the more striking consequences of the above relations.
For simplicity, we shall restrict these remarks to the case of nets
of algebras locally associated with a scalar
Bose field. Since every element 
$\widetilde{\lambda} \in \Ls_+ \setminus \Lid$ of the complement of
the identity component $\Lid$ of the Lorentz group in the
proper Lorentz group $\Ls_+$ can be factored uniquely into a product 
$\widetilde{\lambda} = \theta_R \lambda$, with $\lambda \in \Lid$,
it follows that by defining $U(\widetilde{\lambda}) = J_{W_R} U(\lambda)$
one obtains an (anti-)unitary representation of the
proper Poincar\'e group $\Ps_+$ which acts covariantly upon the original
net of observables. Moreover, denoting by $\Js$ the group generated
by $\{ J_W \mid W \in \Ws\}$ and $\Js_+$ as the subgroup of $\Js$ consisting 
of products of even numbers of the generating involutions 
$\{ J_W \mid W \in \Ws\}$, one has 
\beq \label{lorentz}
\Js = U(\Ps_+) \,\, \textnormal{and} \, \, \Js_+ = U(\Pid) \, .
\eeq
Hence, the modular involutions $\{ J_W \mid W \in \Ws\}$ encode the
isometries of the underlying space--time as well as a representation
of the isometry group which acts covariantly upon the observables. So
in particular, $U(\RR^4) \subset \Js_+$. Recalling that the subgroup
of translations $U(\RR^4)$ determines the dynamics of the quantum
field, one sees that the modular involutions also encode the dynamics
of the model! The dynamics need not be posited, but instead can be
derived from the observables and preparations of the quantum system,
at least in principle, using the modular involutions.
     
     If the quantum field model is such that a scattering theory can
be defined for it and satisfies asymptotic completeness \cite{Ar,Jo,BLT}, 
then the original fields and the asymptotic fields act on the same
Hilbert space and have the same vacuum. Letting $\Rs^{(0)}(W)$, 
$W \in \Ws$, denote the observable algebras associated with the 
free asymptotic field and $J^{(0)}_W$ represent the modular
involution corresponding to $(\Rs^{(0)}(W), \Omega)$, one has,
as was pointed out by B. Schroer \cite{Sch},
$$S = J_{W_R} J_{W_R}^{(0)} \, ,$$
where $S$ is the {\it scattering matrix} for the original field 
model. Hence, the modular involutions associated with the wedge
algebras and the vacuum state also encode all information about the
results of scattering processes in the given model!\footnote{ Note that
the same is {\it not} true about the modular unitaries, since both
the original field and the asymptotic field are covariant under the
same representation of $\Pid$.} 

     In addition, because of the connection between Tomita--Takesaki
modular theory and KMS--states \cite{BrRo2}, modular covariance
entails that when the vacuum state is restricted to $\Rs(W)$ for any
wedge $W$, then with respect to the automorphism group on $\Rs(W)$
generated by the boosts $U(\lambda_W(t))$, it is an equilibrium
state at temperature $1/2 \pi$ (in suitable units). Hence, any
uniformly accelerated observers find when testing the vacuum that it
has a nonzero temperature \cite{Sew}. This striking fact is called the
Unruh effect \cite{Un}. Moreover, because KMS--states are passive
\cite{PuWo}, the vacuum satisfies the second law of thermodynamics
with respect to boosts --- an additional stability property.

     Modular covariance and/or the geometric action of the modular
conjugations (\ref{Jcgma}) have also been derived under other sets of
assumptions (in addition to those discussed in the next section) which
do not refer to the Wightman axioms, \ie purely algebraic settings in
which no appeal to Wightman fields is made \cite{BrGuLo2,Mu,Ku4,Tr}
(see \cite{Bor3} for a review). Thus, these properties and their many
consequences hold quite generally. It is also of interest that some of
these settings provide algebraic versions of the PCT Theorem and the
Spin--Statistics connection \cite{BrGuLo2,GuLo,Mu,Ku1,KuLo}, but we
shall not enter upon this topic here.  We now turn to those conditions
which provide an intrinsic characterization of the vacuum state.

\section{Intrinsic Characterization of the Vacuum State} 
\label{charact}

     Though the definition of a vacuum state given in Definition
\ref{vacuum} is standard, it is not quite satisfactory, since it is
not (operationally) {\it intrinsic}.  It has been seen in Section
\ref{frame} that the elements of quantum theory which are closest to
its operational foundations are states and observables. However, in
the definition of the vacuum state one finds such notions as the
spectrum condition and automorphic (and unitary) representations of
the Poincar\'e group, all of which are not expressed solely in terms
of these states and observables. This may not disturb some readers, so
let us step back and locate the notion of Minkowski space vacuum state
in a larger context.

     One of the primary roles of the vacuum state in quantum field
theory has been to serve as a physically distinguished reference state
with respect to which other physical states can be defined and
referred. Let us recall as an example of this that perturbation theory
is performed with respect to the vacuum state, \ie computations
performed for general states of interest in quantum field theory are
carried out by suitably perturbing the vacuum. This role has proven to
be so central that when theorists tried to formulate quantum field
theory in space--times other than Minkowski space\footnote{After all,
the space--time in which we find ourselves is not Minkowski space.},
they tried to find analogous states in these new settings, thereby
running into some serious conceptual and mathematical problems.  This
is not the place to explain the range and scope of these difficulties,
but one noteworthy problem is indicated by the question: what could
replace the large isometry group (the Poincar\'e group) of Minkowski
space in the definition of ``vacuum state'', in light of the fact that
the isometry group of a generic space--time is trivial? A further
point is that in the definition of ``vacuum state'' the spectrum
condition serves as a stability condition; what could replace it even
in such highly symmetric space--times as de Sitter space, where the
isometry group, though large, does not contain any translations?

     After much effort, a number of interesting selection criteria
have been isolated and studied; see, \eg
\cite{BrFrKo,BS3,BDFS,BFS1,FeVe,KaWa,LuRo,Rad,BoBu,BSads,Ku5,Str}. Of
these, all but one either select an entire folium of states --- \ie
a representation, instead of a state --- or are explicitly limited to
a particular subclass of spacetimes (or both). Here we shall discuss
the selection criterion provided by the Condition of Geometric Modular
Action (CGMA), which in the special case of Minkowski space selects the
vacuum state (as opposed to selecting the entire vacuum
representation) but which can be formulated for general space--times.

     As we now no longer have a vacuum state/representation given, we
return to the notation of Section \ref{frame} and the initial data of
a net $\Os \mapsto \Rs(\Os)$ of observable algebras and a state $\phi$
on $\Rs$. The question we are now examining is: under which
conditions, stated solely in terms of mathematical quantities
completely determined by these initial data, is $\phi$ a vacuum state?
Surprisingly, the core of the answer to this question is the 
relation (\ref{BWcgma}). It will be convenient to introduce the
notation $\Rs_\phi(\Os) \equiv \pi_\phi(\Rs(\Os))'' = (\pi_\phi(\Rs(\Os))')'$.
We consider a special case of the condition first discussed in \cite{BS1}
and subsequently further generalized in \cite{BDFS}.

\begin{definition}
A state $\phi$ on a net $\Os \mapsto \Rs(\Os)$ satisfies the 
\textnormal{Condition of Geometric Modular Action} if the vector
$\Omega_\phi$ is cyclic and separating for $\Rs_\phi(W)$, $W \in \Ws$,
and if the modular conjugation $J_W$ corresponding to 
$(\Rs_\phi(W),\Omega_\phi)$ satisfies
\begin{equation}
J_{W} \, \{ \Rs_\phi(\widetilde{W}) \mid \widetilde{W} \in \Ws \} \, J_{W}  
\subset \{ \Rs_\phi(\widetilde{W}) \mid \widetilde{W} \in \Ws \} \, .
\end{equation}
for all $W \in \Ws$.
\end{definition}

     Note that there is no {\it prima facie} reason why (\ref{BWcgma})
should imply (\ref{Jcgma}). Indeed, why should the action
(\ref{BWcgma}) even be implemented by point transformations on $\RR^4$,
much less by Poincar\'e transformations? And since all Poincar\'e
transformations map wedges to wedges, why should (\ref{Jcgma}) be the
only solution, even if one did find oneself in the latter, fortunate 
situation?

     The following theorem was proven in \cite{BDFS,BS3}. The
interested reader may consult \cite{BS3} for the definition of the
weak technical property referred to in hypothesis (c) of the following
theorem--- a property which involves only the net $W \mapsto \Rs_\phi(W)$
itself.\footnote{In fact, hypothesis (c) may be dispensed with 
if the \msc \, (see below) is satisfied \cite{BS3}.} 

\begin{theorem} [\cite{BDFS,BS3}] \label{cgmathm}
Let $\phi$ be a state on a net $\Os \mapsto \Rs(\Os)$ which satisfies the 
following constraints:

   (a) The map 
$\Ws \ni W \mapsto \Rs_\phi(W) \in \{ \Rs_\phi(W) \mid W \in \Ws \}$ is an 
order-preserving bijection.

   (b) If $W_1 \cap W_2 \neq \emptyset$, then $\Omega_\phi$ is
cyclic and separating for $\Rs_\phi(W_1) \cap \Rs_\phi(W_2)$.
Conversely, if $\Omega_\phi$ is cyclic and separating for
$\Rs_\phi(W_1) \cap \Rs_\phi(W_2)$, then 
$\overline{W_1} \cap \overline{W_2} \neq \emptyset$, where the
bar denotes closure.

   (c) The net $W \mapsto \Rs_\phi(W)$ is locally generated.

   (d) The adjoint action of the modular conjugations $J_W$, $W \in \Ws$,
acts transitively upon the set $\{ \Rs_\phi(W) \mid W \in \Ws \}$, \ie
there exists a wedge $W_0 \in \Ws$ such that
$$\{ J_W \Rs_\phi(W_0) J_W \mid W \in \Ws \} = 
\{ \Rs_\phi(W) \mid W \in \Ws \} \, . $$

     Then there exists a continuous (anti-)unitary representation
$U$ of $\Ps_+$ which leaves $\Omega_\phi$ invariant and acts covariantly
upon the net:
$$U(\lambda) \Rs_\phi(\Os) U(\lambda)^{-1} = \Rs_\phi(\lambda \Os) \, ,$$
for all $\Os$ and $\lambda \in \Ps_+$. Moreover, 
$\Js = U(\Ps_+)$, $\Js_+ = U(\Pid)$ and 
$$J_{W_R} \Rs(\Os) J_{W_R} = \Rs(\theta_R \Os) \, ,$$
for all $\Os$. Furthermore, the wedge duality condition holds:
$$\Rs_\phi(W') = \Rs_\phi(W)' \, ,$$
for all $W \in \Ws$, which entails that the net $W \mapsto \Rs_\phi(W)$
is local. 
\end{theorem}

     Hence, from the state and net are {\it derived} the isometry group of
the space--time; a unitary representation of the isometry group formed
from the modular involutions, leaving the state invariant and
acting covariantly upon the net; the specific geometric action of the
modular involutions found in a special case by Bisognano and Wichmann;
the locality of the net; and even the dynamics \etc. of the theory
(see Section \ref{cgmasec}).

     The conceptually crucial observation is that all conditions in
the hypothesis of this theorem are expressed solely in terms of the
initial net and state, or algebraic quantities completely determined by
them. Condition (a) entails that the adjoint action of the modular
involutions $J_W$ upon the net induces an inclusion preserving
bijection on the set $\Ws$. Condition (b) assures that this bijection
can be implemented by point transformations (indeed Poincar\'e
transformations) \cite{BDFS}, and (c) implies that the representation
$U(\Ps_+)$ is continuous \cite{BS3}.\footnote{Note that the
continuity of the representation of the translation group follows
without condition (c) \cite{BFS}.}  Condition (d)
strengthens the Condition of Geometric Modular Action. Without this
strengthening, the adjoint action of the $J_W$ can still be shown to be
implemented by Poincar\'e transformations \cite{BDFS}, but the group
$\Js$ can then be isomorphic to a proper subgroup of $\Ps_+$ 
\cite{Flodiss}.

     Although such a state $\phi$ is clearly a physically
distinguished state, the spectrum condition and modular covariance
need not be fulfilled \cite{BDFS}.  As an {\it intrinsic} stability
condition, the Modular Stability Condition has been proposed.

\begin{definition} [\cite{BDFS}]
For any $W \in \Ws$, the elements $\Delta_W^{it}$, $t \in \RR$,
of the modular group corresponding to $(\Rs_\phi(W),\Omega_\phi)$
are contained in $\Js$.
\end{definition}

     Note that in this condition no reference is made to the
space--time, its isometry group, or any representation of the isometry
group.  This condition can be posed for models on any space--time
\cite{BDFS}. Together with the CGMA, this modular stability condition
then yields both the spectrum condition and modular covariance
(\ref{modcov}).

\begin{theorem} [\cite{BDFS,BS3}] \label{mscthm}
If, in addition to the hypothesis of Theorem \ref{cgmathm}, the Modular 
Stability Condition is satisfied, then after choosing suitable coordinates
on $\RR^4$, the spectrum condition is satisfied by $U(\Ps_+)$ and
modular covariance holds. The associated representation 
$(\Hs_\phi,\pi_\phi,\Omega_\phi)$ is therefore a vacuum representation.
\end{theorem}

     Of course, this is not, strictly speaking, a characterization of
arbitrary vacuum states; this theorem provides an intrinsic
characterization of those vacuum states which manifest further
desirable properties, properties which are also manifested in the
models in the special circumstances considered by Bisognano and
Wichmann.  But since these latter circumstances are precisely those expected
to arise in standard quantum field theory, the vacuum states
characterized in Theorems \ref{cgmathm} and \ref{mscthm} are probably
the vacuum states of most direct physical interest.

\section{Deriving Space--Time From States and Observables}
\label{derive}

     Although the hypothesis of Theorem \ref{mscthm} makes no explicit
or implicit reference to an underlying space--time, Theorem
\ref{cgmathm} does so implicitly through use of the set of wedges
$\Ws$.\footnote{In fact, only a four dimensional real manifold with a
coordinatization is required in order to formulate and prove the
theorems in Section \ref{charact}, but it is nonetheless clear that
the introduction of wedges as defined tacitly appeals to Minkowski
space.}  However, the results of the preceding section did suggest the
possibility that, without any {\it a priori} reference to a
space--time, the space--time itself as well as an assignment of
localization regions for the observable algebras, along with all of
the above--mentioned results, could be derived from the modular
conjugations associated with a collection of algebras and a suitable
state, as long as the set of modular conjugations verifies certain
purely algebraic relations. And if the Modular Stability Condition is
also satisfied, the state would then be a vacuum state, and the CGMA and
modular covariance would be satisfied.  In fact, this program has been
carried out for a few space--times in \cite{SuWh,SuWh2,SuWh3,Wh}.  In
order to minimize technical complications which would distract
attention away from the essential conceptual point to be made, we will
only discuss this approach in the example of three dimensional
Minkowski space.
          
     To eliminate any reference to a space--time and to strengthen
the purely operational nature of the initial data, we consider a 
collection $\inet$ of unital C$^*$--algebras indexed by ``laboratories''
$i \in I$. $\As_i$ is interpreted as the algebra generated by all
observables measurable in the laboratory $i$.\footnote{The index
set can be naturally refined by further encoding the time (with
respect to some reference clock in the laboratory) during which the
measurement is carried out without changing the validity of the
following assertions.} Since it makes sense
to speak of one laboratory as being contained in another, the
set $I$ of laboratories is provided with a natural partial order
$\leq$. It is then immediate that if $i \leq j$ then
$\As_i \subset \As_j$. Hence, the map $I \ni i \mapsto \As_i \in \inet$ 
is order preserving. We shall assume that this map is a bijection,
since otherwise there would be some redundancy in the description
of the system. If $(I,\leq)$ is a directed set, then $\inet$ is a
net and the inductive limit $\As$ of $\inet$ exists and may be
used as a reference algebra. But even if $\inet$ is not a net,
it is possible \cite{Fred} to naturally embed $\As_i$, $i \in I$,
into a C$^*$--algebra $\As$ so that the inclusion relations are
preserved. It will not be necessary to distinguish between these
cases in the results, and we shall refer to states $\phi$ upon
$\As$ as being states upon the net $\inet$.  

     Given such a state $\phi$, we proceed to the corresponding
GNS representation and define $\Rs_i = \pi_\phi(\As_i)''$,
$i \in I$. We shall assume that the implementing vector $\Omega_\phi$
is cyclic and separating for all $\Rs_i$, $i \in I$, and denote
by $J_i, \Delta_i$, the corresponding modular objects. Again, let
$\Js$ denote the group generated by the involutions $J_i$, $i \in I$.
Note that $J \Omega_\phi = \Omega_\phi$, for all $J \in \Js$.
In this abstract context, the CGMA is the requirement that the
adjoint action of each $J_i$ upon the elements of $\irnet$ leaves
the set $\irnet$ invariant \cite{BDFS}. Among other matters, the
CGMA here entails that the set $\{ J_i \}_{i \in I}$ is an invariant
generating set for the group $\Js$,\footnote{In other words, 
the smallest group containing $\{ J_i \}_{i \in I}$ is $\Js$ and
$J \{ J_i \}_{i \in I} J^{-1} \subset \{ J_i \}_{i \in I}$ for all
$J \in \Js$.} and such a structure is the starting point
for the investigations of the branch of geometry known as
absolute geometry, see {\it e.g.} \cite{Ah,Ba,BBPW}. From such a group
and a suitable set of axioms to be satisfied by the generators of that
group, absolute geometers derive various ``metric'' spaces such as
Minkowski spaces and Euclidean spaces upon which the abstract group $\Js$
now acts as the isometry group of the metric space. Different sets of
axioms on the group yield different metric spaces. This affords us with
the possibility of deriving a space--time from the group $\Js$, 
so that the operational data $(\phi,\inet)$ 
would determine the space--time in which the quantum systems could 
naturally be considered to be evolving. We emphasize that 
different groups $\Js$ would verify different 
sets of algebraic relations and would thus lead to different 
space--times.
 
     For the convenience of the reader, we summarize our standing 
assumptions, which refer solely to objects which are completely
determined by the data $(\phi,\inet)$. 

\medskip

\nind{\bf Standing Assumptions} For the net $\inet$ of nonabelian 
$C^*$-algebras and the state $\phi$ on $\As$ we assume \par
   (i) $i \mapsto \Rs_i$ is an order-preserving bijection; \par
   (ii) $\Omega_\phi$ is cyclic and separating for each algebra 
$\Rs_i$, $i \in I$; \par
   (iii) the adjoint action of each $J_i$ leaves the set $\irnet$ invariant.

\medskip

\nind Already these assumptions restrict significantly the class of
admissible groups $\Js$ \cite{BDFS}. In general, it may be
necessary to pass to a suitable subcollection of $\{ \Rs_i \}_{i \in I}$ 
in order for the Standing Assumptions to be satisfied \cite{BDFS} 
(if, indeed, they are satisfied at all) --- see \cite{SuWh} for a brief 
discussion of this point.

     We must introduce some notation in order to concisely formulate
the algebraic requirements upon $\Js$ which lead to the construction
of three dimensional Minkowski space. We use lower case Latin letters to 
denote arbitrary modular involutions $J_i$, $i \in I$, upper case Latin 
letters to denote involutions in $\Js$ of the form $ab$, and lower case 
Greek letters for arbitrary elements of $\Js$. By $\xi \mid \eta$ we shall 
mean ``$\xi\eta$ is an involution'', and $\alpha, \beta \mid \xi, \eta$ is 
shorthand for ``$\alpha \mid \xi$, $\beta \mid \xi$, $\alpha \mid \eta$, and
$\beta \mid \eta$''.

\begin{theorem} [\cite{SuWh}] 
Assume in the above setting that the following relations hold in $\Js$: 

1. For every $P,Q$ there exists a $g$ with $P,Q \mid g$.

2. If $P,Q \mid g,h$, then $P = Q$ or $g = h$.

3. If $a,b,c \mid P$, then $abc \in \{ J_i : i \in I \}$.

4. If $a,b,c \mid g$, then $abc \in \{ J_i : i \in I \}$.

5. There exist $g,h,j$ such that $g \mid h$ but 
$j \mid g,h,gh$ are all false.

6. For each $P$ and $g$ with $P \mid g$ false, there exist exactly
two distinct elements $h_1,h_2$ such that $h_1,h_2 \mid P$ is true and
$g,h_i \mid R,c$ are false for all $R,c$, $i = 1,2$.

     Then there exists a model (based on $\Js$)
of three dimensional Minkowski space in which each 
$J_i$, $i \in I$, is identified as a spacelike line (and every 
spacelike line is such an element) and on which each $J_i$, $i \in I$, 
acts adjointly as the reflection about the spacelike
line to which it corresponds. $\Js$ is isomorphic to $\Ps_+$\footnote{the
proper Poincar\'e group for three dimensional Minkowski space} and forms in
a canonical manner a strongly continuous (anti)unitary representation
$U$ of $\Ps_+$. Moreover, there exists a bijection
$\chi : I \rightarrow \Ws$\footnote{the set of wedges in 
three dimensional Minkowski space} such that after defining
$\Rs(\chi(i)) = \Rs_i$, the resultant net $\{ \Rs(\chi(i)) \}$
of wedge algebras on Minkowski space is covariant under the action
of the representation $U(\Ps_+)$. Furthermore, one has
$\Rs(\chi(i))' = \Rs(\chi(i)')$ for all $i \in I$. Thus, if the
map $\chi : I \rightarrow \Ws$ is order--preserving, then the
net $\{ \Rs(\chi(i)) \}$ is local.

     If, further, $\Delta_j^{it} \in \Js$ for all $j \in I$, $t \in \RR$,
then modular covariance is satisfied and the state $\omega$ is a vacuum
state on the net $\{ \Rs(\chi(i)) \}$.
\end{theorem}

     We emphasize that assumptions 1--6 are purely algebraic in nature
and involve only the group $\Js$, which is completely determined by
the initial data $(\phi,\inet)$. Although we do not propose the
verification of such conditions as a practical procedure to determine
space--time, it is, in our view, a noteworthy conceptual point that
such a derivation is possible in principle. It is also noteworthy that
the derived structure is so rigid and provides such a complete basis
for physical interpretation. Indeed, from the observables and state
can be derived a space--time, an identification of the localizations
of the observables in that space--time and a continuous unitary
representation of the isometry group of the space--time such that the
resultant, re--interpreted net is covariant under the action of the
isometry group and the re--interpreted state is a vacuum state. It is
perhaps worth mentioning that the modular symmetry group $\Js$ of a
theory on four dimensional Minkowski space as discussed in Section
\ref{charact} {\it does not} verify assumptions 1--6 above.  Moreover,
models on three dimensional Minkowski space satisfying the CGMA do
verify assumptions 1--6.

     In \cite{SuWh2,SuWh3,Wh} sets of algebraic conditions on $\Js$
have also been found so that the space derived is three dimensional
de Sitter space, respectively four dimensional Minkowski space.
We anticipate that similar results can be proven for other highly
symmetric space--times such as anti--de Sitter space and the 
Einstein universe, but not for general space--times.

\section{Concluding Remarks} \label{concl}

     It is a striking fact that, in the senses indicated above, the 
modular involutions associated with the vacuum state (and {\it only}
the vacuum state) encode the following physically significant
matters.

\begin{itemize}
\item the space--time in which the quantum systems may be viewed as evolving
\item the isometry group of the space--time
\item a strongly continuous unitary representation of this isometry group
which acts covariantly upon the net of observable algebras and leaves
the state invariant
\item the locality, \ie the Einstein causality, of the quantum systems
\item the dynamics of the quantum systems
\item the scattering behavior of the quantum systems
\item the spin--statistics connection in the quantum systems
\item the stability of the quantum systems
\item the thermodynamic behavior of the quantum systems
\end{itemize}

     It has also become clear through examples --- quantum field
theories on de Sitter space \cite{BoBu,BDFS,Flodiss}, anti-de Sitter space
\cite{BFS1,BSads}, a class of positively curved Robertson--Walker
space--times \cite{BMS1,BMS2}, as well as others \cite{SuVe,Str} --- that
the encoding of crucial physical information by modular objects and
the subsequent utility of this approach are not limited to Minkowski space
theories.

     It is necessary to distinguish between the, in some sense,
maximal results of Section \ref{derive} and those of Section
\ref{charact}. The former cannot be expected to be reproducible in
most space--times, since the isometry groups are not large enough to
determine the space--time, and the arguments in Section \ref{derive}
rely tacitly upon the possibility of interpreting the modular group
$\Js$ as (a suitably large subgroup of) the isometry group of some
space--time.  However, most of the results of Section \ref{charact},
and hence most of the list above, can be expected to be attainable in
more general space--times, without regard to the size of the isometry
group of the space--time. As has been verified in a class of models in
a family of Robertson--Walker space--times \cite{BMS1}, the CGMA and
the encoding of crucial physical information by modular involutions
associated with certain observable algebras and select states can hold
even when the modular symmetry group $\Js$ is strictly larger than the
isometry group of the space--time (in fact, in these examples a
significant portion of $\Js$ is not associated with any kind of
pointlike transformations upon the space--time).  In other words, it
is quite possible that the fact that the modular symmetry group gives
no more than (a subgroup of) the isometry group of the space--time in
the presence of the CGMA for theories on Minkowski or de Sitter space
is an accident due to the fact that these space--times are maximally
symmetric. Moreover, it is possible that using the CGMA and Modular
Stability Condition to select states of physical interest yields a
modular symmetry group $\Js$ containing, along with the standard
symmetries expected from classical theory, new and purely quantum
symmetries encoding unexpected physical information (further evidence
for this speculation which goes beyond \cite{BMS1} can be adduced in
\cite{FaSch}).

     Finally, we mention that modular objects associated with
privileged algebras of observables and states (usually the vacuum) are
also proving to be useful in the construction of quantum field models
in two, three and four dimensional Minkowski space, which cannot be
constructed by previously known techniques of constructive quantum
field theory \cite{BrGuLo3,Sch2,BuLe,MuSchYng,BuSu2,Le3,GrLe,BS6}. But
such matters go well beyond the scope of this paper.

\end{document}